\documentclass[a4paper,11pt]{article}
\usepackage{jinstpub} 
\usepackage{lineno}
\usepackage{siunitx}
\usepackage{booktabs}
\usepackage{tabularx}
\usepackage{array}

\title{\boldmath All-directional gamma-ray imaging using a $\mathrm{NaI(Tl)}$ scintillator with double-sided SiPM readout}

\author[a,b]{Anzori Sh. Georgadze} 
\affiliation[a]{ Institute for Nuclear Research of the National Academy of Sciences of Ukraine,\\Prospekt Nauky 47, 03680, Kyiv, Ukraine}
\affiliation[b]{Advanced Sciences OÜ,\\Jaama tn, Tartu, Estonia}    
\emailAdd{a.sh.georgadze@gmail.com}

\abstract{
Gamma-ray imaging systems capable of determining the direction of incident radiation are essential for homeland security, nuclear non-proliferation, environmental monitoring, and radiological emergency response. Conventional directional detectors typically rely on heavy mechanical collimators, coded apertures, or complex multi-layer Compton cameras, resulting in significant trade-offs between detection efficiency, system weight, and complexity.

This work presents a compact, high-efficiency omnidirectional gamma-ray imaging concept based on a monolithic cylindrical $2\times2$\,inch NaI(Tl) scintillation crystal coupled to dual-ended $16\times16$ Silicon Photomultiplier (SiPM) matrices. The system exploits scintillation light distributions collected from both crystal faces to reconstruct three-dimensional interaction positions and depth-of-interaction (DOI).

Detailed GEANT4 Monte Carlo simulations ($4.0 \times 10^8$ primary histories) incorporating full optical photon transport were performed for 662\,keV gamma rays from a $^{137}$Cs source. The simulated energy resolution is $6.69\% \pm 0.31\%$ FWHM at the photopeak. A hybrid directional reconstruction framework is implemented, combining volumetric self-attenuation (active masking) for robust low-energy localization with intra-crystal Compton imaging for higher energies. 

With approximately 40,000 accumulated photopeak counts, the active-masking algorithm achieves angular resolutions of FWHM$_{\rm elev} \approx 5.7^\circ$ and FWHM$_{\rm az} \approx 3.7^\circ$. The system fully complies with the EN IEC 62327 standard for handheld radionuclide identification devices. Under the required 120-second acquisition window, it suppresses terrestrial background (NORM) from the lower hemisphere by a factor of $\approx$320, improving to $\approx$980 with 300-second integration.

These results demonstrate that a monolithic dual-ended NaI(Tl) detector can transform a conventional scalar spectrometer into a sensitive, real-time directional imaging instrument suitable for portable field use and automated cargo inspection.
}


\arxivnumber{1234.56789} 

\begin{document}
\maketitle
\flushbottom

\section{Introduction}
\label{sec:intro}

Gamma-ray imaging systems capable of determining the direction of incident radiation are essential in numerous fields, including homeland security, nuclear safeguards, environmental monitoring, astrophysics, and medical imaging \cite{knoll2010, parajuli2022development}. Unlike conventional detectors that provide only count rate or spectroscopic information, directional systems enable source localization and radiation-field mapping in complex environments \cite{cooper2023networked, Tian2026OptimizationOA}.

Current gamma-ray imaging technologies rely primarily on mechanical collimation, coded apertures, or Compton scattering reconstruction \cite{Smith2001Hybrid, Susaiev2023New}. Each approach involves fundamental trade-offs between angular resolution, detection efficiency, field of view, and system complexity. Mechanically collimated and coded-aperture systems typically suffer from low efficiency or heavy shielding requirements, while conventional Compton cameras often require multi-layer segmented detectors, leading to increased complexity, dead space, and parallax effects \cite{todd1974, vetter2019compton}.

Hybrid approaches combining active masking with Compton imaging have recently gained attention as a means to overcome these limitations \cite{peterson2009, chivers2018}. Active masking exploits self-attenuation within the detector volume to provide directional sensitivity at lower energies, while Compton kinematics extends performance at higher energies. This dual-mode strategy offers the potential for wide dynamic range within a single compact detector.

In this work, we investigate an omnidirectional gamma-ray imaging concept based on a monolithic cylindrical NaI(Tl) scintillator with dual-ended Silicon Photomultiplier (SiPM) matrix readout. The cylindrical geometry provides near-$4\pi$ coverage while maintaining mechanical simplicity. Simultaneous readout from opposing crystal faces enables reconstruction of both transverse position and depth-of-interaction (DOI) through analysis of scintillation light distributions.

The detector performance is evaluated using detailed GEANT4 Monte Carlo simulations incorporating full optical photon transport. A $2''\times2''$ NaI(Tl) crystal coupled to $16\times16$ SiPM arrays on both ends is modeled and irradiated with 662\,keV gamma rays from a $^{137}$Cs source. Particular attention is given to compliance with the international standard EN IEC 62327 for handheld radionuclide identification devices.

This paper presents the detector concept, simulation methodology, reconstruction algorithms (active masking and Compton back-projection), and quantitative directional performance results. The proposed architecture demonstrates the feasibility of compact, high-efficiency directional gamma-ray detection without mechanical collimation.

\section{Materials and methods}
\label{sec:methods}

\subsection{GEANT4 modeling of a NaI(Tl) scintillation detector}
\label{sec:geant4}

\begin{figure}[b]
\centering
\includegraphics[width=0.55\linewidth]{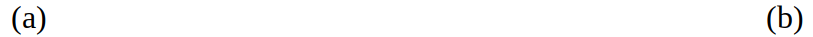}
\includegraphics[width=0.55\linewidth]{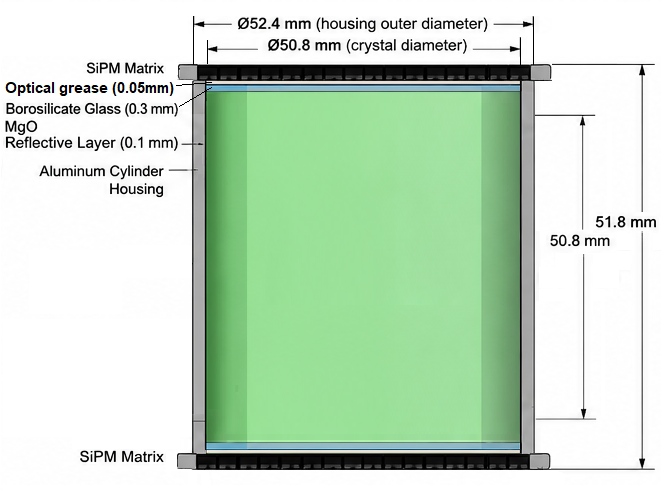}
\includegraphics[width=0.38\linewidth]{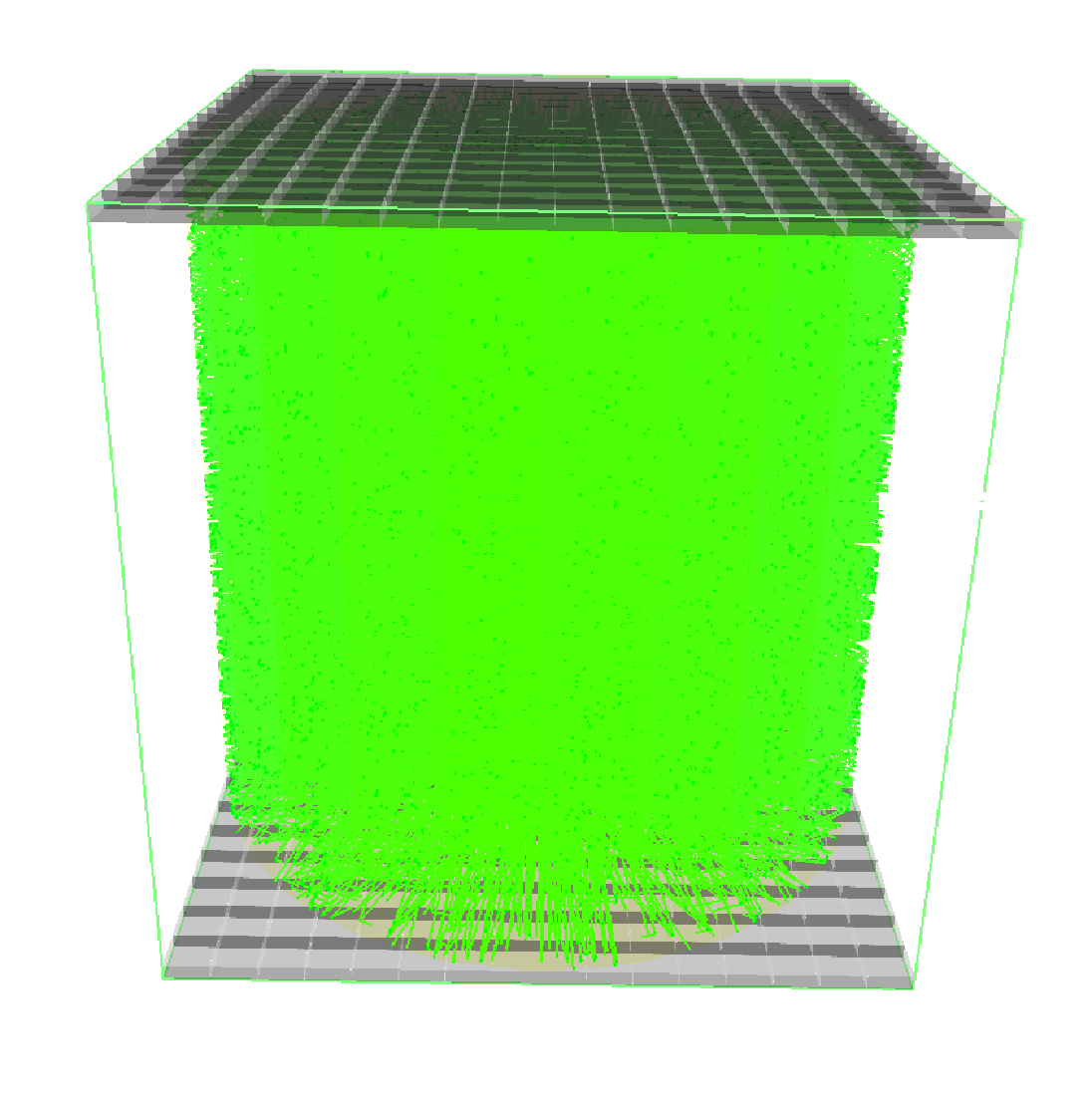}
\caption{(a) Schematic of the dual-ended readout NaI(Tl) detector showing the coordinate system and SiPM array placement. (b) Example GEANT4 visualization of a 662\,keV $^{137}$Cs interaction. Light green lines represent optical photon trajectories propagating toward the top and bottom SiPM arrays. The resulting light distribution on each array encodes the three-dimensional position of the interaction.}
\label{fig:figure1}
\end{figure}

The detector consists of a monolithic cylindrical NaI(Tl) crystal with dimensions $\varnothing 2$\,inch $\times$ 2\,inch ($50.8\,\mathrm{mm} \times 50.8\,\mathrm{mm}$) as shown in figure~\ref{fig:figure1}. The crystal is housed in a light-tight aluminum enclosure internally coated with MgO powder (reflectivity $\approx 97\%$). Both end faces are sealed with borosilicate glass windows, and $16\times16$ SiPM arrays are optically coupled to each window using optical grease. This configuration yields 256 channels per face and 512 channels in total.

The detector response was modeled using the GEANT4 Monte Carlo simulation toolkit~\cite{geant4}, incorporating full electromagnetic, optical photon transport, and boundary physics. The simulated geometry consists of a $2\times2$\,inch cylindrical NaI(Tl) crystal enclosed in an aluminum housing. The top and bottom faces are coupled to borosilicate glass windows, while the lateral surfaces are lined with MgO reflective material.

Key optical parameters for NaI(Tl) at the emission peak ($\lambda=415$\,nm) include: density $3.67$\,g/cm³, scintillation yield $38{,}000$ photons/MeV, decay time $230$\,ns, and refractive index $n=1.85$, bulk absorption length 50~cm. Optical photons are transported using the \texttt{G4OpticalPhysics} module with the GLISUR model for boundary processes~\cite{collaboration2019book, gumplinger2002optical}. The MgO reflector is modeled as a diffuse (\texttt{groundbackpainted}) surface with reflectivity 0.98, and the aluminum housing as a polished surface with reflectivity 0.88. Optical coupling between the crystal, glass window, and SiPM is realized via a thin layer of optical grease ($n=1.47$).

A Hamamatsu S13360-series SiPM was used as a reference photodetector. The effective photon detection efficiency (PDE), weighted by the NaI(Tl) emission spectrum, was set to approximately 40\%.

Electromagnetic interactions were simulated using the \texttt{G4EmLivermorePhysics} list. A total of $4.0 \times 10^8$ primary $^{137}$Cs gamma-ray histories were generated to accumulate sufficient statistics in the photopeak and Compton continuum. For each event, the integrated charge in all 512 SiPM channels (256 top + 256 bottom) was recorded and served as input to the subsequent energy- and position-reconstruction algorithms.

\section{Gamma source direction determination}
\label{sec:direction}

\subsection{Compton imaging via intra-crystal two-interaction reconstruction}
\label{sec:compton}

A key innovation of the proposed architecture is its ability to perform Compton imaging entirely within a single monolithic NaI(Tl) crystal. This is accomplished by identifying and reconstructing two-interaction sequences — a primary Compton scatter followed by photoelectric absorption — using high-granularity SiPM readout.

\begin{figure}[t]
\centering
\includegraphics[width=0.52\linewidth]{figures/ab.png}
\includegraphics[width=0.75\linewidth]{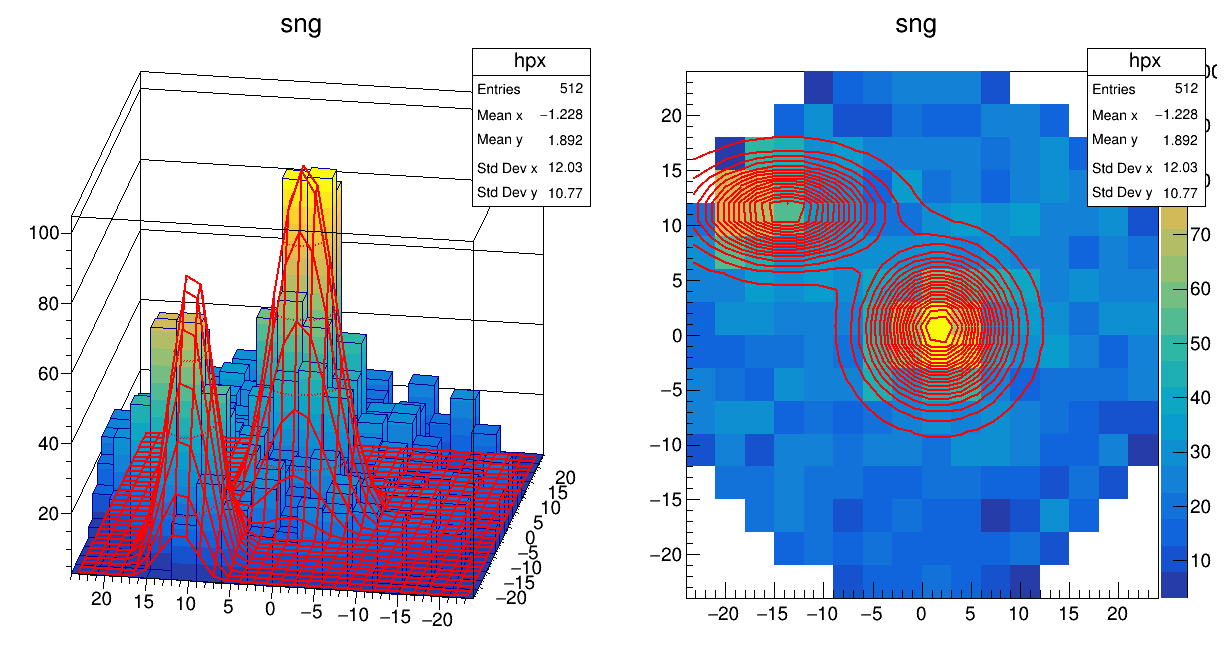}
\caption{Example of a multi-interaction event in the monolithic scintillator. (a) 3D light distribution. (b) 2D projection on the SiPM plane. A two-component Gaussian mixture model is used to resolve the positions and relative energies of the two interaction sites.}
\label{fig:gmm}
\end{figure}

For a valid Compton event, the two interaction sites produce distinct scintillation light pools. These overlapping distributions are decomposed using a two-component Gaussian-mixture model (or non-linear least-squares fit of light response functions) applied to the summed top+bottom SiPM response (Figure~\ref{fig:gmm}).

The Compton scattering angle $\theta_C$ is calculated from the deposited energies $E_1$ and $E_2$ using the kinematic relation:
\begin{equation}
\cos\theta_C = 1 - m_ec^2\left(\frac{1}{E_2} - \frac{1}{E_1+E_2}\right),
\label{eq:compton_basic}
\end{equation}
where $m_ec^2 \approx 511$\,keV. The resulting Compton cone has its apex at the first (scatter) interaction position, with the opening angle $\theta_C$.

Because the dual-ended readout does not provide sufficient timing resolution to determine the chronological order of the two interactions, both possible orderings are initially considered. This leads to a two-fold ambiguity that manifests as a secondary mirror lobe in the back-projection map. The ambiguity is partially mitigated by using the independent active-masking direction estimate (Section~\ref{sec:activemask}) as a coarse prior to select the more kinematically consistent ordering.

\paragraph{Interaction Position Reconstruction.}

The full reconstruction pipeline consists of four stages: (1) 3D position and energy reconstruction of each interaction site, (2) determination of the correct interaction sequence, (3) construction of the event-wise Compton cone, and (4) accumulation of many cones into a sky map.

The measured light distribution on the SiPM arrays is modeled as a superposition of single-interaction light response functions (LRFs):
\begin{equation}
Q_{\mathrm{top,bottom}}(u,v) \approx \sum_{i=1}^{N} E_i \, L_{\mathrm{top,bottom}}(u,v\,|\,x_i,y_i,z_i).
\label{eq:lrf}
\end{equation}

Candidate interaction sites are first identified via peak finding on the summed charge map. Refined 3D positions $(x_i,y_i,z_i)$ and deposited energies $E_i$ are obtained through non-linear least-squares fitting, with initial depth estimates derived from the localized top-to-bottom asymmetry:
\begin{equation}
R_i = \frac{Q_i^{\rm top} - Q_i^{\rm bottom}}{Q_i^{\rm top} + Q_i^{\rm bottom}}.
\label{eq:doi_compton}
\end{equation}

\paragraph{Interaction sequencing and cone construction.}

The most probable scattering sequence is selected by comparing the kinematic scattering angle (from energies) with the geometric angle (using the active-masking direction as a prior). Accepted events are used to construct Compton cones, which are then back-projected onto the sky to form a directionality map.

Even with realistic 3\,mm position smearing and finite energy resolution, the superposition of cones from multiple events produces a statistically significant peak at the true source direction, albeit with broader angular resolution ($\sim$44$^\circ$ FWHM at 662\,keV) compared to the active masking approach.

\subsection{Determination of gamma-ray source direction using active masking}
\label{sec:activemask}

Single-source directional accuracy is quantified by the mean absolute error and FWHM of the residual distribution between true and reconstructed source positions. The proposed detector exploits an active masking mechanism based on self-absorption and self-shielding of scintillation light within the monolithic crystal volume. Scintillation photons originating closer to a given photodetector face experience shorter optical paths and lower attenuation, resulting in position-dependent light collection across the dual SiPM arrays.

\begin{figure}[b]
\centering
\includegraphics[width=0.55\linewidth]{figures/ab.png}
\includegraphics[width=0.45\linewidth]{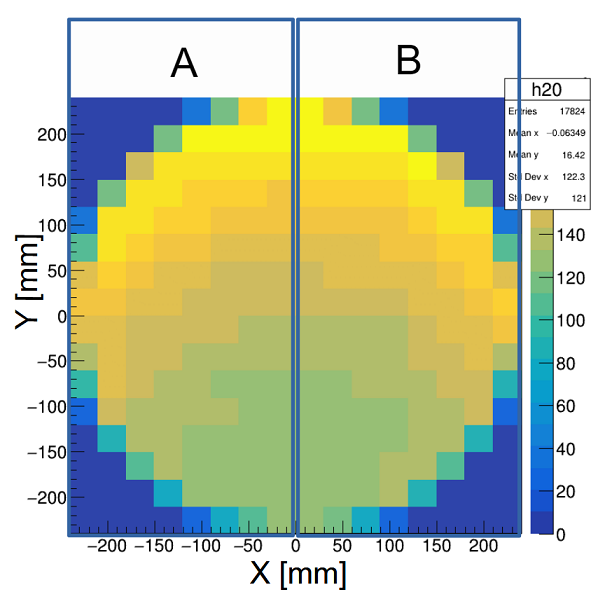}
\includegraphics[width=0.50\linewidth]{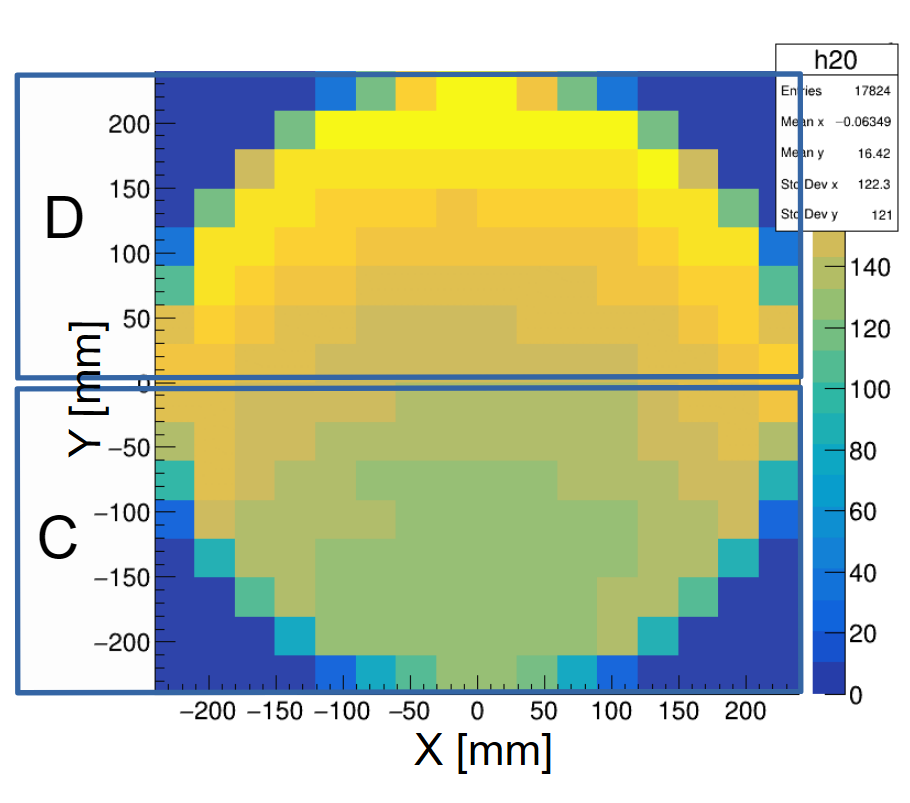}
\caption{Definition of the four overlapping edge sectors ($A$--$D$) on the summed dual $16\times16$ SiPM response. Each sector covers an $8\times16$ pixel region. Sectors $A$/$B$ and $C$/$D$ provide sensitivity to horizontal and vertical asymmetries, respectively.}
\label{fig:sectors}
\end{figure}

The summed charge map from both SiPM arrays is defined as:
\begin{equation}
M_{i,k} = T_{i,k} + B_{i,k},
\label{eq:summed_map}
\end{equation}
where $T_{i,k}$ and $B_{i,k}$ are the calibrated charges from the top and bottom arrays. A central exclusion band (rows 4--7) is applied to suppress isotropic internal scattering and enhance peripheral directional contrast:
\begin{equation}
M^{*}_{i,k} =
\begin{cases}
0, & \text{if } 4 \le i \le 7, \\
M_{i,k}, & \text{otherwise}.
\end{cases}
\label{eq:mask_operator}
\end{equation}

Four overlapping edge sectors ($A$--$D$) are then integrated to extract asymmetry vectors (show in figure~\ref{fig:sectors}):
\begin{align}
Q_A &= \sum_{i=0}^{15}\sum_{k=0}^{3} M^{*}_{i,k}, &
Q_B &= \sum_{i=0}^{15}\sum_{k=12}^{15} M^{*}_{i,k}, \label{eq:sigmaAB} \\
Q_C &= \sum_{i=0}^{3}\sum_{k=0}^{15} M^{*}_{i,k}, &
Q_D &= \sum_{i=12}^{15}\sum_{k=0}^{15} M^{*}_{i,k}. \label{eq:sigmaCD}
\end{align}

\subsubsection{Azimuthal Angle Reconstruction}
\label{sec:azimuth}

The azimuthal source angle is determined from the orthogonal asymmetry vectors:
\begin{align}
d_x &= Q_B - Q_A, \label{eq:dx} \\
d_y &= Q_D - Q_C. \label{eq:dy}
\end{align}

The final azimuth is computed using the four-quadrant inverse tangent:
\begin{equation}
\theta_{\rm source} = \arctan2(d_y, d_x) \cdot \frac{180^\circ}{\pi}.
\label{eq:azimuth_final}
\end{equation}
This approach provides robust azimuthal reconstruction by exploiting the geometric self-masking effect of the cylindrical crystal, without the need for external collimation.

\subsubsection{Elevation angle reconstruction}
\label{sec:elevation}

The source elevation angle is reconstructed by exploiting the dual-ended SiPM configuration, which simultaneously collects scintillation light from opposing ends of the cylindrical crystal. The total charge collected by the top and bottom photodetector arrays is defined as:
\begin{equation}
Q_{\mathrm{top}} = \sum_{i=0}^{15}\sum_{k=0}^{15} T_{i,k}, \qquad
Q_{\mathrm{bottom}} = \sum_{i=0}^{15}\sum_{k=0}^{15} B_{i,k},
\label{eq:sigmatopbot}
\end{equation}
where $T_{i,k}$ and $B_{i,k}$ are the calibrated charge signals from the top and bottom arrays, respectively.

The normalized charge asymmetry along the longitudinal (z) axis is then calculated as:
\begin{equation}
\alpha_z = \frac{Q_{\mathrm{top}} - Q_{\mathrm{bottom}}}{Q_{\mathrm{top}} + Q_{\mathrm{bottom}}}, \qquad \alpha_z \in [-1,1].
\label{eq:alphaz}
\end{equation}
This parameter serves as a sensitive depth-of-interaction (DOI) metric.

The source elevation angle $\theta_{\rm elevation}$ (ranging from $-90^\circ$ to $+90^\circ$) is obtained via a calibration function:
\begin{equation}
\theta_{\rm elevation} = f(\alpha_z).
\label{eq:elevation_final}
\end{equation}
A first-order approximation $f(\alpha_z) \approx \arcsin(\alpha_z)$ provides a simple analytical baseline. For high accuracy across the full range, an empirical look-up table is constructed from dedicated GEANT4 calibration simulations using point sources at known elevations ($0^\circ$ to $45^\circ$ in $11.25^\circ$ steps). Linear interpolation is used between calibration points.

\begin{figure}[t]
\centering
\includegraphics[width=0.72\linewidth]{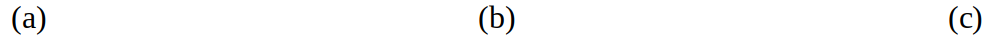}
\includegraphics[width=0.65\linewidth]{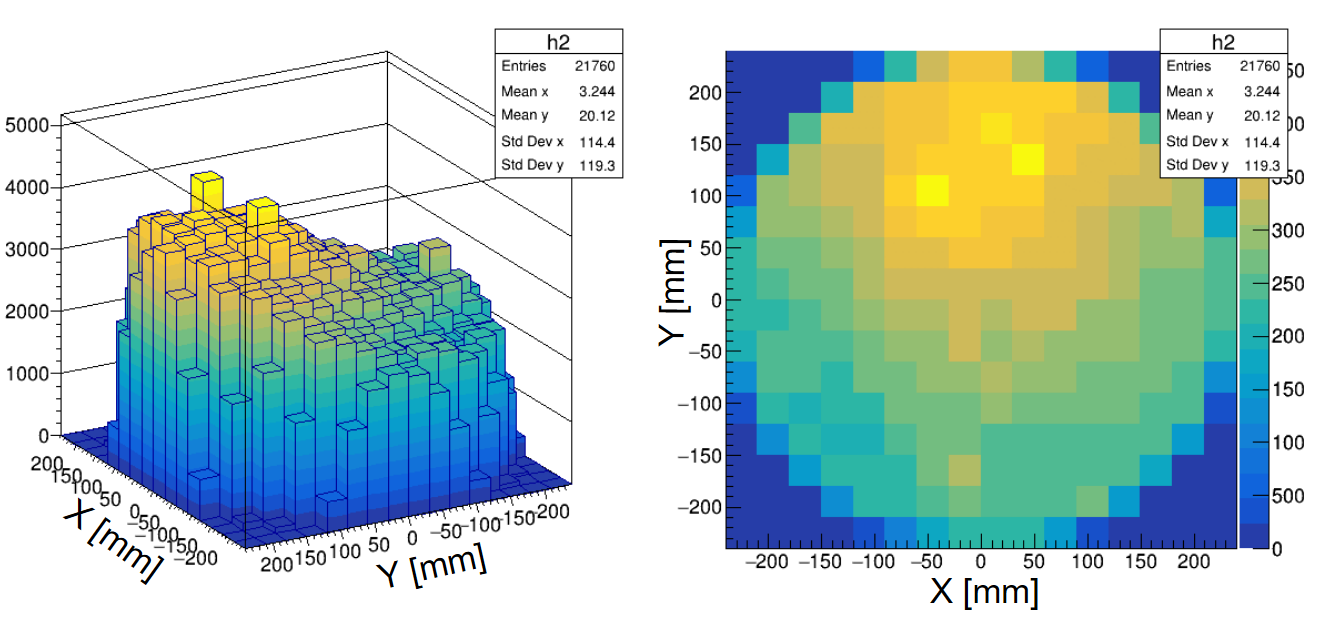}
\includegraphics[width=0.33\linewidth]{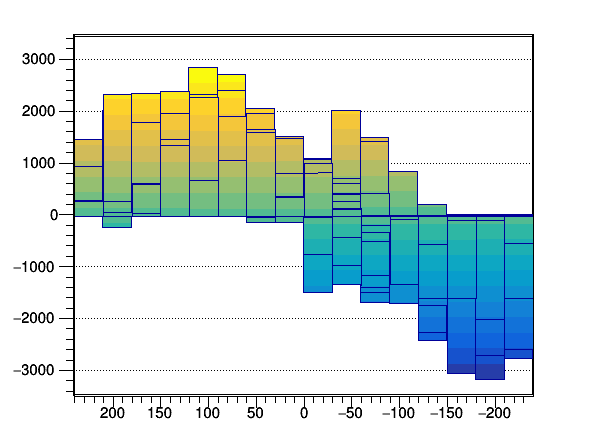}
\caption{(a) Three-dimensional map of the summed response $Q_{\rm sum} = Q_{\rm top} + Q_{\rm bottom}$ for $2.17\times10^4$ detected 662\,keV events, showing spatially uniform total light collection. (b) Differential response $Q_{\rm diff} = Q_{\rm top} - Q_{\rm bottom}$, where positive (negative) values indicate dominant light collection at the top (bottom) face. (c) One-dimensional projection of $Q_{\rm diff}$ along the horizontal axis.}
\label{fig:figure4}
\end{figure}

The differential response maps in Figure~\ref{fig:figure4}(b,c) clearly reveal the axial asymmetry induced by source elevation. When the source is displaced vertically relative to the detector center, the cylindrical geometry causes a systematic change in the effective material thickness traversed by incoming gamma rays. This self-attenuation effect shifts the mean depth-of-interaction toward the forward-facing side of the crystal, producing a reproducible relationship between source elevation and the measured charge asymmetry $\alpha_z$.

By combining this elevation estimate with the independently determined azimuthal angle $\phi_{\rm azimuth}$, the system reconstructs a full 3D direction vector $(\phi, \theta)$ to the source. This approach enables near-$4\pi$ directional sensitivity using a compact monolithic detector, without requiring mechanical collimators, detector rotation, or complex multi-module arrays.

\section{Results}
\label{sec:results}

\subsection{Gamma-ray spectroscopy}
\label{sec:spectroscopy}

\begin{figure}[b]
  \centering
\includegraphics[width=0.62\linewidth]{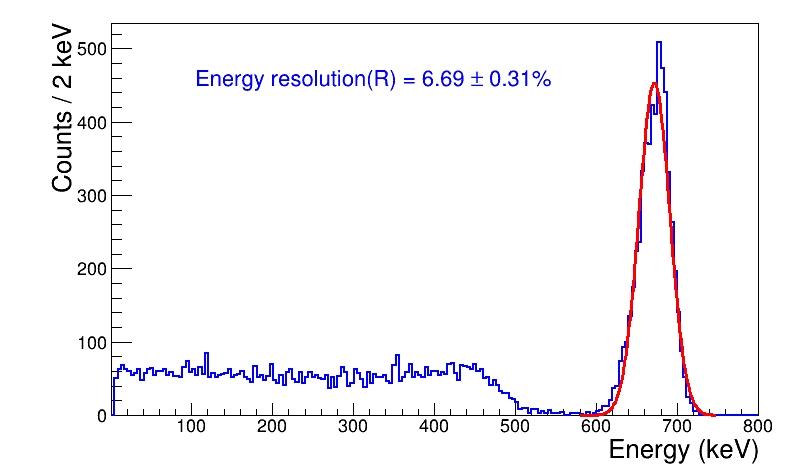}  
  \caption{Simulated pulse-height distribution of a $^{137}$Cs source   modeled via GEANT4 for a cylindrical $2\times2$ inch NaI(Tl) detector.   A Gaussian fit applied to the full-energy photopeak yields an energy   resolution of $6.69\%\pm0.31\%$ FWHM at 662~keV. The continuous Compton   continuum and its corresponding edge at approximately 477~keV are  cleanly resolved, validating the physical transport models.}
  \label{fig:spectrum}
\end{figure}

The pulse-height spectrum for a $^{137}$Cs gamma-ray source, simulated using the GEANT4 Monte Carlo toolkit to benchmark the light-transport and energy-deposition parameters of the cylindrical NaI(Tl) detector, is shown in Fig.~\ref{fig:spectrum}. The resulting pulse-height distribution is dominated by a symmetrically broadened full-energy photopeak centered at
661.7~keV and a continuous Compton-scattering baseline bounded sharply by the kinematic Compton edge at 477.3~keV.

A non-linear least-squares Gaussian fit applied directly to the simulated full-energy peak reveals a FWHM of 44.3~keV at the peak centroid, corresponding to an intrinsic energy resolution of $R=6.69\%\pm0.31\%$. This performance metric is in good agreement with the typical
experimental range (6.5\%--7.5\%) reported for physical laboratory-grade NaI(Tl) crystals under equivalent geometries~\cite{knoll2010}, cross-validating the optical simulation parameters used throughout this study.

\subsection{Compton imaging via intra-crystal two-interaction reconstruction}
\label{sec:compton_results}

Compton imaging reconstructs the direction to a gamma-ray source by measuring the positions and deposited energies of two successive interactions within the monolithic scintillator: an initial Compton scatter followed by photoelectric absorption. The reconstructed interaction coordinates and energies $(E_1, E_2)$ define a cone on the sky that contains the possible source direction. Superposition of many such cones allows statistical localization of the source.

In this analysis, transverse positions $(X_1,Y_1)$ and $(X_2,Y_2)$ of the two interaction vertices are determined from a two-component Gaussian-mixture fit to the summed SiPM response. Depths of interaction $(Z_1, Z_2)$ are estimated using the per-cluster top-to-bottom charge asymmetry (Eq.~\ref{eq:doi_compton}). Because the dual-ended readout does not provide time-of-flight information to resolve the chronological order of the two interactions, \emph{both} possible orderings are back-projected for each event. The resulting two-fold ambiguity produces a characteristic mirror lobe in the reconstructed sky map.

\begin{figure}[b]
\centering
\includegraphics[width=\linewidth]{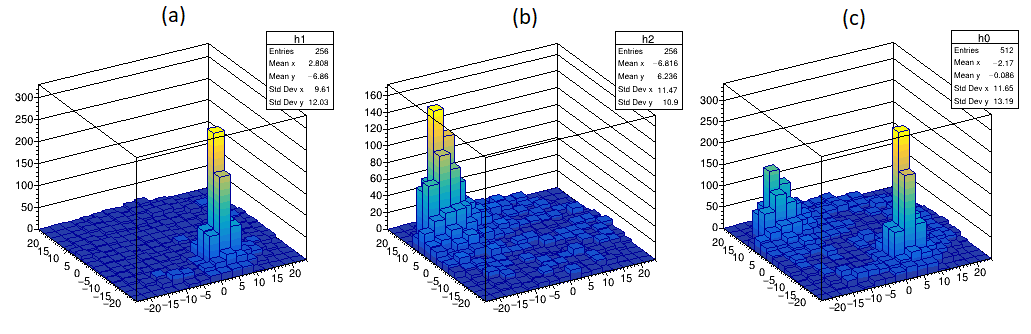}
\caption{Simulated scintillation light profiles for a multi-interaction event. The panels show (left) the top SiPM array, (center) the bottom SiPM array, and (right) the channel-by-channel sum. Summing the arrays significantly improves the separation of the two interaction sites.}
\label{fig:multisite}
\end{figure}

To improve vertex separation, the light distributions from the top and bottom SiPM arrays are summed channel-by-channel:
\begin{equation}
Q_{\mathrm{sum},i} = Q_{\mathrm{top},i} + Q_{\mathrm{bottom},i}.
\label{eq:qsum}
\end{equation}
This combined 2D profile yields higher contrast and better resolves closely spaced interactions compared to using individual arrays (Figure~\ref{fig:multisite}).

Event selection proceeds in two stages. First, the summed charge map $Q_{\rm sum}$ is searched for exactly two spatial maxima using \texttt{TSpectrum2} (ROOT). Events with fewer or more peaks are discarded. The detected peak coordinates initialize a two-component Gaussian-mixture fit, yielding refined transverse positions. Depths are then calculated from the localized asymmetry ratios.

A kinematic filter is subsequently applied to retain only physically valid two-interaction events within the $^{137}$Cs photopeak. For an incident energy $E_0 = 661.7$\,keV, the kinematic limits for Compton scattering are $E_{e,\max} \approx 478$\,keV (maximum recoil electron energy at $180^\circ$) and $E'_{\min} \approx 184$\,keV (minimum scattered photon energy). The final selection criteria are:
Total deposited energy: $620 \le E_1 + E_2 \le 700$\,keV; Lower deposited energy: $50 < E_{\rm low} \le 256$\,keV; Higher deposited energy: $E_{\rm high} \ge 420$\,keV.

These tightened windows provide margin against finite energy resolution (6.69\% FWHM) while suppressing noise, partial energy depositions, and random coincidences.

\paragraph{Compton back-projection image reconstruction and event selection}

\begin{figure}[b]
\centering
\includegraphics[width=0.55\linewidth]{figures/ab.png}
\includegraphics[width=0.65\linewidth]{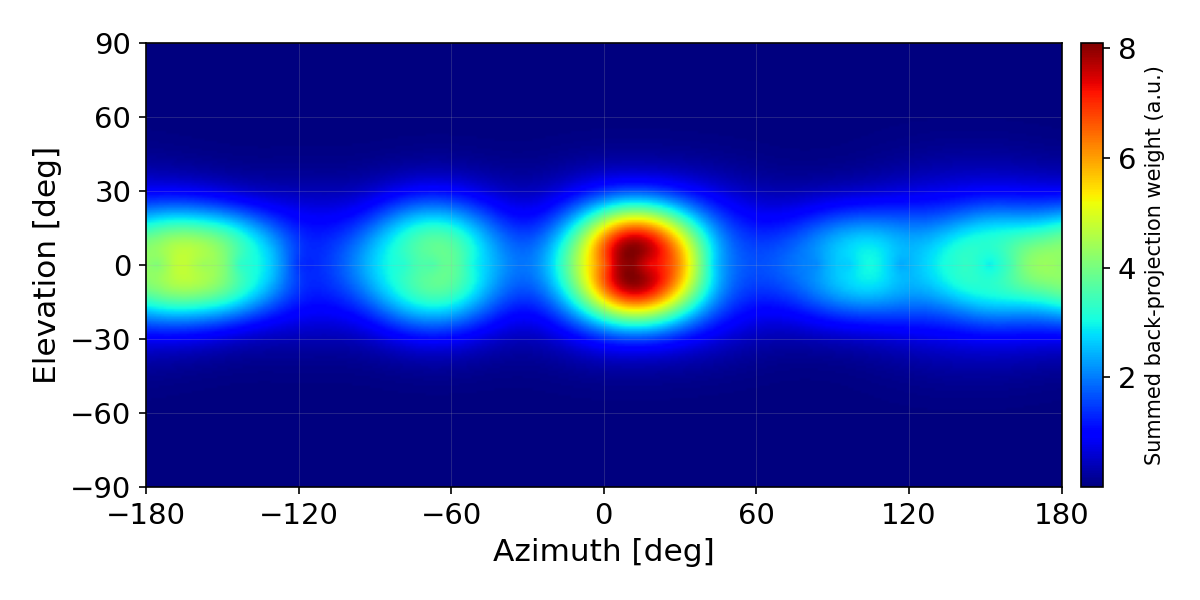}
\includegraphics[width=0.32\linewidth]{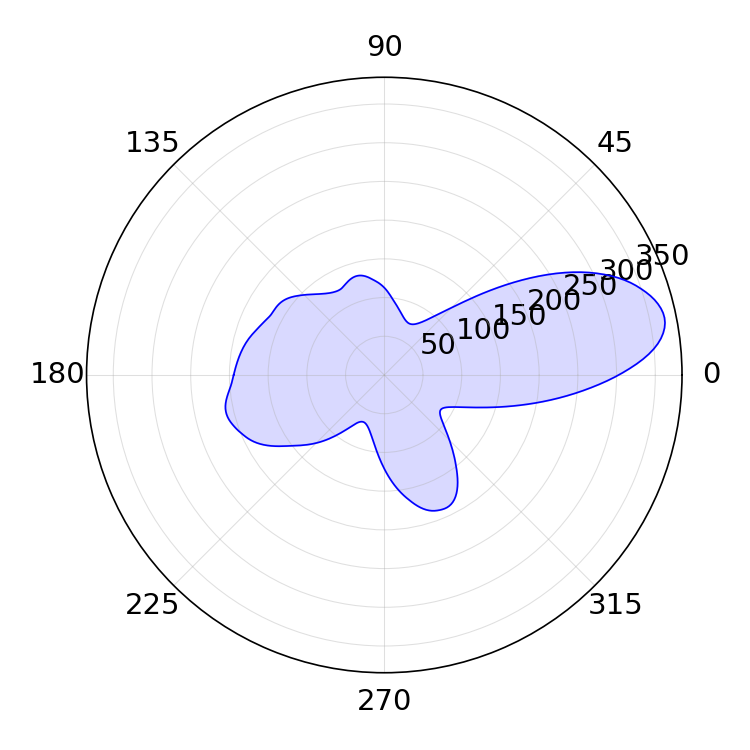}
\caption{(a) Reconstructed Compton back-projection sky map in elevation ($-90^\circ$ to $+90^\circ$) and azimuth ($-180^\circ$ to $+180^\circ$) for 130 filtered events with 3\,mm spatial smearing. (b) Integrated azimuthal distribution showing the primary lobe near $0^\circ$ and a symmetric mirror lobe near $295^\circ$ caused by sequencing ambiguity.}
\label{fig:skymap}
\end{figure}

The spatial distribution of reconstructed source directions obtained via continuous Compton back-projection is presented in figure~\ref{fig:skymap}(a). To ensure physical fidelity and maximize signal-to-noise ratio, a stringent kinematic and topological filtering protocol was applied. From approximately 20,000 simulated gamma-ray events falling within the photopeak window, only about 200 events with two clearly resolved interaction sites were initially selected using a multi-cluster peak-search algorithm. Enforcing a minimum 15\,mm spatial separation between interaction vertices further reduced this population to 130 high-quality double-interaction events ($\approx 0.65\%$ of total photopeak statistics).

To emulate realistic detector performance, the true interaction coordinates were stochastically smeared with a 3\,mm Gaussian ($\sigma_{\rm pos} = 3$\,mm), corresponding to the effective position uncertainty arising from SiPM pixel size, internal light propagation, and diffuse reflections within the MgO-reflective housing. This degradation highlights the inherent difficulty of resolving multiple interactions in a compact monolithic crystal, where scintillation light pools from nearby vertices significantly overlap.

Additionally, the dual-ended SiPM readout does not provide sufficient charge-sharing asymmetry to unambiguously determine the chronological order of interactions. Consequently, both possible sequencing permutations were retained for each event, resulting in two back-projected cones per history. This two-fold ambiguity produces a characteristic dual-lobe structure in the sky map.
The angular uncertainty of individual Compton cones is governed by the propagation of the detector’s energy resolution (6.69\% FWHM at 662\,keV) and the 3\,mm position uncertainty through the kinematic Compton formula. This results in broad, smeared back-projection arcs. When the full imaging space is projected onto a 360$^\circ$ azimuthal plot (Figure~\ref{fig:skymap}(b)), an angular resolution of \textbf{FWHM$_{\rm Compton} \approx 44^\circ$} ($\sigma \approx 18.7^\circ$) is achived.

\subsection{Directional performance using active masking}
\label{sub:active_masking}

The directional capability of the detector was evaluated through GEANT4 simulations involving 130 discrete point-like $^{137}$Cs sources (662\,keV). These sources were distributed over five elevation angles ($0^\circ$, $11.25^\circ$, $22.5^\circ$, $33.75^\circ$, $45^\circ$) and 32 azimuthal angles ($0^\circ$ to $348.75^\circ$ in $11.25^\circ$ steps). This setup provided comprehensive hemispherical coverage to characterize the volumetric self-occlusion (active masking) effect of the cylindrical NaI(Tl) crystal.

For a 2-minute integration time ($\sim$10,000 net photopeak events), the elevation error distribution is characterized by $\mu = -0.23^\circ$ and $\sigma = 3.75^\circ$, corresponding to a full width at half maximum (FWHM) of \textbf{8.8$^\circ$}. The azimuthal error shows $\mu = 0.44^\circ$, $\sigma = 3.09^\circ$, and FWHM = \textbf{7.3$^\circ$}.

Extending the acquisition to 5 minutes ($\sim$40,000 net photopeak events) significantly improves the light distribution statistics. The elevation error tightens to $\mu = -0.04^\circ$, $\sigma = 2.41^\circ$, and FWHM = \textbf{5.7$^\circ$}, while the azimuthal resolution improves to $\mu = 0.14^\circ$, $\sigma = 1.56^\circ$, and FWHM = \textbf{3.7$^\circ$}.

\begin{figure}[htbp]
    \centering
    \includegraphics[width=1.0\textwidth]{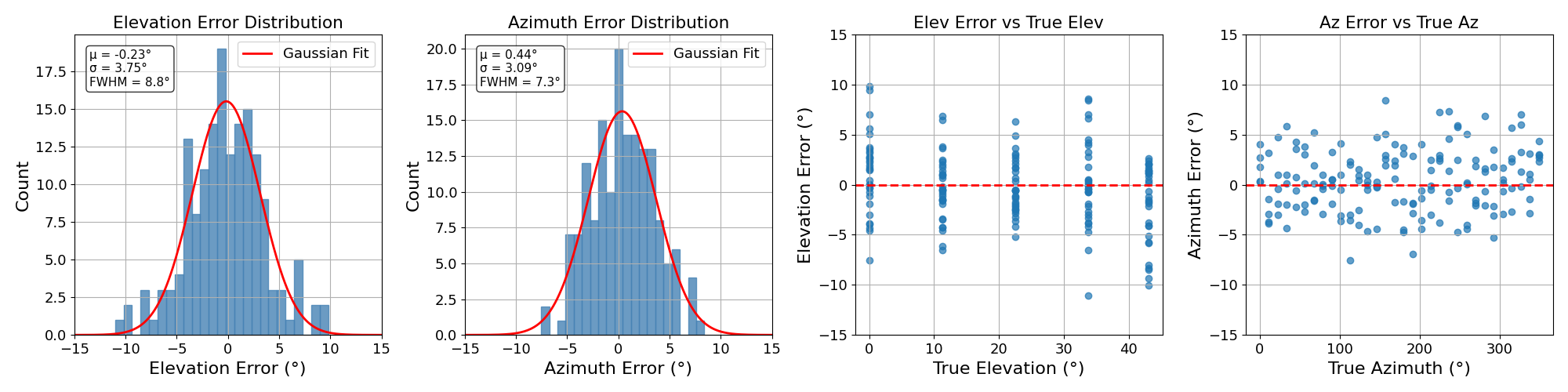}
    \includegraphics[width=1.0\textwidth]{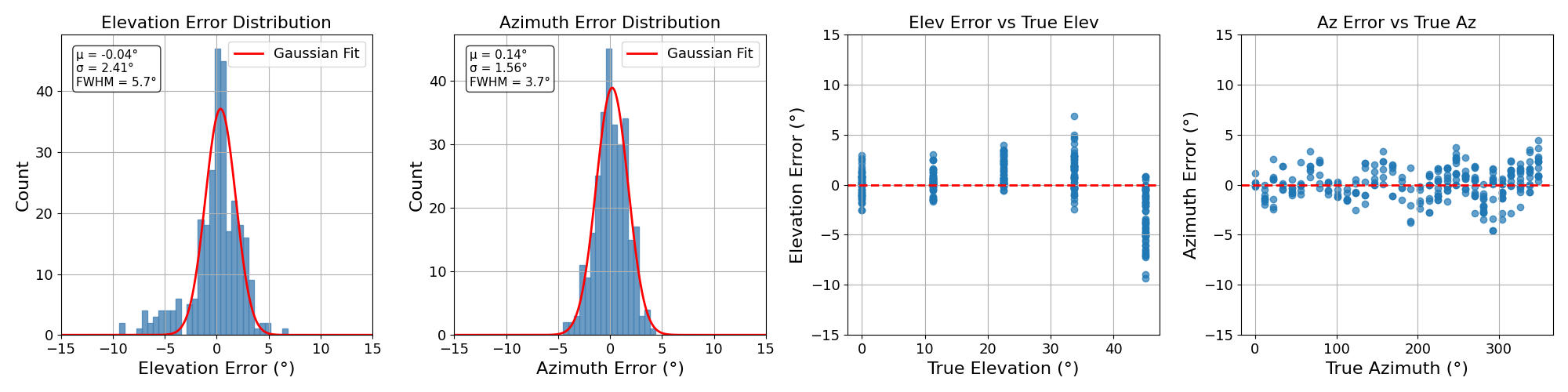}
    \caption{Angular reconstruction performance for a $^{137}$Cs source (662\,keV) at 1\,m distance. \textit{Top panels:} Results for a 2-minute acquisition ($\sim$10,000 photopeak counts). \textit{Bottom panels:} Results for a 5-minute acquisition ($\sim$40,000 photopeak counts). Left columns show error histograms with Gaussian fits; right columns display error versus true angle. The system achieves FWHM values of $8.8^\circ$ (elevation) / $7.3^\circ$ (azimuth) at 2 minutes, improving to $5.7^\circ$ / $3.7^\circ$ at 5 minutes.}
    \label{fig:figure8}
\end{figure}

The scatter plots of reconstruction error versus true angle (figure~\ref{fig:figure8}, bottom panels) confirm that the method is essentially unbiased across the entire tested range. These results demonstrate that the dual-ended SiPM readout, combined with K-Means clustering of SiPM hits and localized top-to-bottom charge asymmetry, enables robust three-dimensional interaction position reconstruction. Consequently, the intrinsic self-attenuation of a standard $2\times2$\,inch NaI(Tl) crystal can be effectively harnessed as an active mask, providing meaningful directional information without mechanical collimators or external shielding.

\begin{figure}[t]
    \centering
    \includegraphics[width=0.5\textwidth]{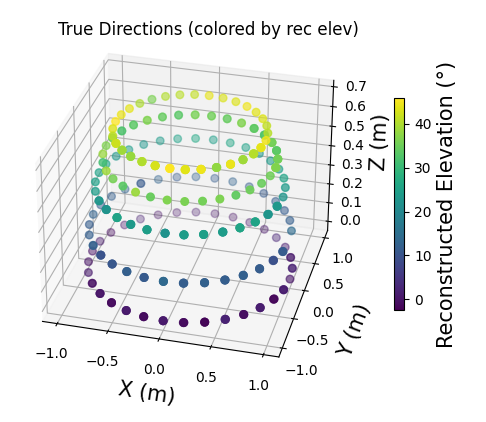}
    \caption{Three-dimensional representation of true source directions on a unit sphere (1\,m radius) for the 2-minute acquisition. Each point is color-coded by the reconstructed elevation angle. The smooth, monotonic color progression across latitude rings demonstrates accurate depth-of-interaction estimation using top-to-bottom charge asymmetry.
    }
    \label{fig:figure9}
\end{figure}

Figure~\ref{fig:figure9} shows the three-dimensional reconstruction of true source directions on a unit sphere, with each point color-coded by the reconstructed elevation angle.

\section{Discussion}
\label{sec:discussion}

The GEANT4 simulation results demonstrate that the proposed dual-ended SiPM readout NaI(Tl) scintillation detector provides significant angular sensitivity to incident gamma radiation. The reconstructed response exhibits well-defined angular distributions in both azimuthal and polar coordinates, with effective angular resolutions on the order of a few degrees (see Table~\ref{tab:resolution_summary}). This capability enables the decomposition of the radiation field into narrow directional components, in contrast to conventional scintillation-based radioisotope identification devices (RIIDs), which treat the detector response as essentially isotropic.

A key finding of this study is that the active masking approach (based on volumetric self-attenuation and localized charge asymmetry) substantially outperforms Compton kinematics reconstruction at 662\,keV. While the active masking method produces a sharp, geometrically constrained Point Spread Function (PSF) determined by the physical boundaries of the crystal, Compton back-projection suffers from large per-event angular uncertainty, boundary scattering effects, and two-fold degeneracy (mirror solutions). Consequently, active masking dominates the directional performance at medium energies, whereas Compton kinematics is expected to become more competitive at higher energies ($>2$\,MeV), where forward scattering is strongly favored and energy resolution improves.

\begin{table}[t]
  \centering
  \caption{Summary of directional resolution (FWHM) for the active-masking reconstruction algorithm under different photopeak statistics conditions.}
  \label{tab:resolution_summary}
  \vspace{2mm}
  \begin{tabular}{lccc}
    \hline
    Condition & Photopeak counts & FWHM$_{\mathrm{elev}}$ & FWHM$_{\mathrm{az}}$ \\
    \hline
    Low-statistics (Compton back-projection) & $\approx 130$ & -- & $44.0^\circ$ \\
    EN IEC 62327 test scenario (120\,s) & $\approx 10{,}200$ & $8.8^\circ$ & $7.3^\circ$ \\
    Extended statistics (300\,s) & $\approx 40{,}000$ & $5.7^\circ$ & $3.7^\circ$ \\
    \hline
  \end{tabular}
\end{table}

The strong dependence of angular resolution on accumulated statistics highlights the importance of acquisition time in field applications. Even under the strict 120-second limit imposed by international standards, the system achieves directional performance that is highly useful for source localization. Longer integration times yield further improvements, demonstrating excellent scalability. These results confirm that a compact monolithic NaI(Tl) detector with dual-ended readout can transition from purely spectroscopic operation to true directional radiation detection without mechanical collimators or significant increases in system complexity.
\paragraph{Relevance and Compliance with EN\,IEC\,62327}

To assess the operational readiness of the proposed dual-sided SiPM readout architecture for regulatory deployment, its performance was evaluated against the international standard \textbf{EN IEC 62327} \textit{(Hand-held instruments for the detection and identification of radionuclides)} \cite{eniec623272019}.

The EN IEC 62327 standard specifies that a commercial Radionuclide Identification Device (RID) must successfully detect and identify a target radionuclide under a net ambient dose equivalent rate increase of $\dot{H}^*_{10} = 0.5\,\mu\mathrm{Sv/h}$ at the detector face, within a maximum automated analysis time of $\tau_{\mathrm{test}} = 120\,\mathrm{s}$.

For a laboratory-grade $2'' \times 2''$ cylindrical NaI(Tl) crystal, a $^{137}$Cs source (662\,keV) producing a net dose rate of $0.5\,\mu\mathrm{Sv/h}$ yields a photopeak count rate of approximately 80--90 counts/s. Over the 120-second integration window required by the standard, this results in the accumulation of a statistically significant number of net full-energy events:
\begin{equation}
N_{\mathrm{photo,net}} = R_{\mathrm{photo}} \times \tau_{\mathrm{test}} 
\approx 85\,\mathrm{counts/s} \times 120\,\mathrm{s} \approx 10{,}200\,\mathrm{counts}.
\label{eq:nphoto}
\end{equation}

Simulation results demonstrate that with this count population (120-second integration), the system achieves an angular resolution of approximately \textbf{8.8$^\circ$} (FWHM) in elevation and \textbf{7.3$^\circ$} (FWHM) in azimuth. This performance confirms that the proposed detector not only fully complies with the spectroscopic identification requirements of EN IEC 62327, but additionally provides reliable directional information within the same regulatory time frame.

\paragraph{Terrestrial background reduction and minimum detectable activity}

Border-crossing points and cargo inspection facilities impose demanding operational conditions: short acquisition times, high throughput, and challenging background environments. Conventional NaI(Tl)-based RIIDs and portal monitors integrate radiation over a large solid angle and struggle to discriminate weak localized sources from spatially heterogeneous terrestrial background.

Unlike cosmic-ray or airborne backgrounds, terrestrial natural radiation (primarily $^{40}$K and the $^{238}$U/$^{232}$Th decay chains in soil, concrete, and infrastructure) is predominantly concentrated in the lower hemisphere. Therefore, the effective background solid angle for an uncollimated detector is better approximated as $\Omega_{\rm bg} = 2\pi \approx 6.28\,\mathrm{sr}$ rather than the full $4\pi$.

The proposed directional system defines a reduced effective solid angle estimated as
$\Delta\Omega \approx \Delta\phi \times \sin(\theta_{\rm center}) \, \Delta\theta$  (expressed in steradians, with angles converted to radians).
This yields a significant background rejection capability. Under the EN IEC 62327 120-second protocol, the achieved angular resolution of $8.8^\circ \times 7.3^\circ$ corresponds to:
\begin{equation}
\Delta\Omega_{120\mathrm{s}} \approx 0.0196\,\mathrm{sr}, \qquad
\mathrm{BRR}_{120\mathrm{s}} = \frac{2\pi}{\Delta\Omega_{120\mathrm{s}}} \approx \mathbf{321}.
\end{equation}

When the acquisition time is extended to 300 seconds, the improved resolution of $5.7^\circ \times 3.7^\circ$ reduces the effective solid angle to $\Delta\Omega_{300\mathrm{s}} \approx 0.00642\,\mathrm{sr}$, increasing the background rejection ratio to:
\begin{equation}
\mathrm{BRR}_{300\mathrm{s}} = \frac{2\pi}{\Delta\Omega_{300\mathrm{s}}} \approx \mathbf{979}.
\end{equation}

This substantial reduction in accepted background directly improves the minimum detectable activity (MDA) according to the Currie formula:
\begin{equation}
\mathrm{MDA} = \frac{2.71 + 4.65\sqrt{B}}{\varepsilon P_\gamma t},
\label{eq:mda}
\end{equation}
where $B$ is the background counts in the region of interest, $\varepsilon$ is the full-energy peak efficiency, $P_\gamma$ is the gamma emission probability, and $t$ is the acquisition time.

When the background is reduced by a factor $f = \Delta\Omega / (2\pi)$, the directional MDA improves approximately as
\begin{equation}
\mathrm{MDA}_{\rm dir} \approx \mathrm{MDA}_{\rm iso} \sqrt{f}.
\end{equation}
Thus, background rejection factors of 321 (120\,s) and 979 (300\,s) translate into roughly 18-fold and 31-fold improvements in sensitivity, respectively, compared to a conventional uncollimated detector.

\section{Conclusions}
\label{sec:conclusions}

In this work, a novel all-directional gamma-ray tracking methodology was developed and evaluated using a compact monolithic $2\times2$\,inch NaI(Tl) scintillation crystal read out by dual-ended $16\times16$ SiPM matrices. High-fidelity GEANT4 Monte Carlo simulations, including detailed optical photon transport, were employed to model the detector response. The simulated pulse height spectrum for a $^{137}$Cs source yielded an energy resolution of $6.69\% \pm 0.31\%$ FWHM at 662\,keV, in excellent agreement with typical experimental values for laboratory-grade NaI(Tl) crystals and validating the simulation framework.

The study demonstrates that the proposed dual-ended readout architecture enables effective directional reconstruction through a combination of active masking (volumetric self-attenuation) and 3D interaction position reconstruction based on K-Means clustering of SiPM hits and top-to-bottom charge asymmetry. Active masking proved particularly effective at 662\,keV, achieving angular resolutions of \textbf{7.3$^\circ$} (FWHM, azimuth) and \textbf{8.8$^\circ$} (FWHM, elevation) within the 120-second integration time mandated by the EN IEC 62327 standard. With extended 5-minute acquisitions, these values improve to \textbf{3.7$^\circ$} and \textbf{5.7$^\circ$}, respectively.

While Compton kinematics reconstruction shows lower accuracy at 662\,keV due to limited scattering angles and energy resolution, it is expected to become highly competitive at multi-MeV energies (e.g., 4.4\,MeV from $^{12}$C and 6.13\,MeV from $^{16}$O in proton therapy), where the differential cross-section is strongly forward-peaked and larger energy depositions improve angular precision. This establishes the proposed detector as a wide-dynamic-range system, with active masking dominating at lower energies and Compton imaging contributing at higher energies.

The system fully complies with the EN IEC 62327 standard for radionuclide identification. Under the required $0.5\,\mu\mathrm{Sv/h}$ net dose rate increase, the detector accumulates over 10,000 net photopeak counts in 120\,s. Within this regulatory timeframe, it provides real-time directional localization while suppressing terrestrial background (NORM) from the lower hemisphere by a factor of approximately \textbf{320}. Extending the measurement to 5 minutes increases this rejection ratio to nearly \textbf{980}, significantly enhancing sensitivity.

Overall, this work establishes that precise gamma-ray source localization is achievable with a compact, monolithic scintillation detector without mechanical collimation. The combination of spectroscopic identification, directional capability, and strong background rejection makes the proposed architecture highly promising for applications in border security, cargo inspection, nuclear safeguards, environmental monitoring, and medical physics.

Future work will focus on experimental validation with physical prototypes, optimization under realistic background conditions, extension to a broader energy range, and the integration of machine-learning techniques to further improve multi-interaction reconstruction.

\acknowledgments

This research did not receive any specific grant from funding agencies. 

\bibliographystyle{JHEP}
\bibliography{biblio.bib}

\providecommand{\href}[2]{#2}\begingroup\raggedright\begin{thebibliography}{10}

\bibitem{knoll2010}
G.F.~Knoll, \emph{Radiation Detection and Measurement}, John Wiley \& Sons, Hoboken, NJ, 4th~ed. (2010).

\bibitem{parajuli2022development}
R.K.~Parajuli, M.~Sakai, R.~Parajuli and M.~Tashiro, \emph{Development and applications of compton camera—a review}, \href{https://doi.org/https://doi.org/10.3390/s22197374}{\emph{Sensors} {\bfseries 22} (2022) 7374}.

\bibitem{cooper2023networked}
R.~Cooper, N.~Abgrall, G.~Aversano, M.~Bandstra, D.~Hellfeld, T.~Joshi et~al., \emph{Networked sensing for radiation detection, localization, and tracking}, \href{https://doi.org/https://doi.org/10.1088/1742-6596/2586/1/012125}{\emph{Journal of Physics: Conference Series} {\bfseries 2586} (2023) }.

\bibitem{Tian2026OptimizationOA}
C.-S.~Tian, J.~Yang, G.~Zeng, X.-Y.~Yang, H.~Deng, C.-H.~Hu et~al., \emph{Optimization of a dual-end readout bar-shaped scintillator detector for compton imaging}, \href{https://doi.org/https://doi.org/10.1007/s41365-026-01964-9}{\emph{Nuclear Science and Techniques} {\bfseries 37} (2026) }.

\bibitem{Smith2001Hybrid}
L.E.~Smith, C.Y.~Chen, D.K.~Wehe and Z.~He, \emph{Hybrid collimation for industrial gamma-ray imaging: Combining spatially coded and compton aperture data}, \href{https://doi.org/10.1016/S0168-9002(00)01148-7}{\emph{Nuclear Instruments \& Methods in Physics Research Section A-accelerators Spectrometers Detectors and Associated Equipment} {\bfseries 462} (2001) 576}.

\bibitem{Susaiev2023New}
Y.~Susaiev, V.~Schoepff and O.~Limousin, \emph{New 3d coded aperture collimator for x/gamma-ray wide-field imaging},  in \emph{EPJ Web of Conferences}, 2023, \href{https://doi.org/10.1051/epjconf/202328807011}{DOI}.

\bibitem{todd1974}
R.W.~Todd, J.M.~Nightingale and D.B.~Everett, \emph{A proposed $\gamma$-camera}, \href{https://doi.org/10.1038/251132a0}{\emph{Nature} {\bfseries 251} (1974) 132}.

\bibitem{vetter2019compton}
K.~Vetter, R.~Barnowski and A.~Haefner, \emph{Advances in {Compton} imaging for nuclear security and environmental monitoring}, \href{https://doi.org/10.1016/j.ppnp.2019.02.001}{\emph{Progress in Particle and Nuclear Physics} {\bfseries 106} (2019) 35}.

\bibitem{peterson2009}
T.E.~Peterson et~al., \emph{Hybrid coded aperture and {Compton} imaging using an active mask}, \href{https://doi.org/10.1016/j.nima.2009.06.043}{\emph{Nuclear Instruments and Methods in Physics Research Section A: Accelerators, Spectrometers, Detectors and Associated Equipment} {\bfseries 604} (2009) 114}.

\bibitem{chivers2018}
D.H.~Chivers et~al., \emph{Transparent ceramic garnet gamma-ray spectrometer with directionality}, \href{https://doi.org/10.1109/TNS.2018.2850438}{\emph{IEEE Transactions on Nuclear Science} {\bfseries 65} (2018) 2199}.

\bibitem{geant4}
{\scshape GEANT4 Collaboration} collaboration, \emph{{GEANT4}---a simulation toolkit}, \href{https://doi.org/10.1016/S0168-9002(03)01368-8}{\emph{Nuclear Instruments and Methods in Physics Research Section A: Accelerators, Spectrometers, Detectors and Associated Equipment} {\bfseries 506} (2003) 250}.

\bibitem{collaboration2019book}
{Geant4 Collaboration}, ``{Book for application developers}.'' \url{https://geant4-userdoc.web.cern.ch/UsersGuides/ForApplicationDeveloper/fo/BookForApplicationDevelopers.pdf}, 2023.

\bibitem{gumplinger2002optical}
P.~Gumplinger, ``Optical photon processes in geant4. users’ workshop at cern, november 2002. availableonline:.'' \url{https://geant4-internal.web.cern.ch/sites/default/files/geant4/support/training/users_workshop_2002/lectures/OpticalPhoton.pdf}.

\bibitem{eniec623272019}
{European Committee for Electrotechnical Standardization (CENELEC)}, \emph{Radiation protection instrumentation - hand-held instruments for the detection and identification of radionuclides and for the estimation of ambient dose equivalent rate from photon radiation},  Standard \href{https://standards.iteh.ai/catalog/standards/clc/4d032df4-b9ca-4e04-aeae-f791fa0424d5/en-iec-62327-2019}{EN IEC 62327:2019}, European Committee for Electrotechnical Standardization (CENELEC), Brussels, BE (2019).

\end{thebibliography}\endgroup

\end{document}